\documentclass[aps,pra,twocolumn,superscriptaddress]{revtex4}
\usepackage{amsfonts}
\usepackage{amssymb}
\usepackage{amsmath}
\usepackage{epsfig}
\usepackage{color}
\usepackage{graphics, graphicx}
\usepackage{bbold}
\usepackage{psfrag}
\usepackage{mathcomp}
\usepackage{subfigure}
\usepackage{verbatim}
\usepackage[colorlinks,citecolor=blue]{hyperref}

\begin{document}


\title{Readout of Majorana bound states via Landau-Zener transition}


\author{Zhen-Tao Zhang}
\email[]{zhangzhentao@lcu.edu.cn}
\affiliation{School of Physics Science and Information Technology, Shandong Key Laboratory of Optical Communication Science and Technology, Liaocheng University, Liaocheng 252059, China}

\author{Dong E. Liu}
\affiliation{State Key Laboratory of Low Dimensional Quantum Physics, Department of Physics, Tsinghua University, Beijing, 100084, China}
\affiliation{Beijing Academy of Quantum Information Sciences, Beijing 100193, China}
\affiliation{Frontier Science Center for Quantum Information, Beijing 100184, China}


\date{\today}

\begin{abstract}
Reading out Majorana bound states (MBSs) is essential both to verify their non-Abelian property and to realize topological quantum computation. Here, we construct a protocol to measure the parity of two MBSs in a Majorana island coupled to double quantum dot (DQD). The parity information is mapped to the charge state of the DQD through Landau-Zener transition. The operation needed is sweeping the bias of the DQD, which is followed by charge sensing. In the case without fine-tuning, a single run of sweep-and-detection implement a weak measurement of the parity. We find that in general a sequence of about ten runs would completely project a superposition state to either parity, and the charge detection in each run records how the state of MBSs collapses step by step. Remarkably, this readout protocol is of non-demolition and robust to low frequency charge fluctuation.
 
\end{abstract}


\maketitle

\section{Introduction}
Majorana bound states (MBSs) are expected to obey the non-Abelian statistics and hold the promise to realize topologically-protected quantum computation \cite{Kitaev01,Ivanov01,Nayak08,Akhmerov10}. Among the numerous candidate systems that may host MBSs, one-dimensional semiconductor nanowire proximited with superconductor is widely studied \cite{Lutchyn10,Oreg10}, and considered as the most promising platform for topological quantum computing. Although many remarkable transport properties of this system are consistent with MBSs \cite{Mourik12,Das12,Deng12,Rokhinson12,Churchill13,Finck13,Albrecht16,Deng16,Chen17,Suominen17,Nichele17,Onder18}, the novel non-Abelian braiding statistics of MBSs are not confirmed yet. To this end, one need to 
achieve two types of key operations in experiment: braiding a pair of MBSs and measuring the fermion parity of Majorana pairs. These operations are also lying at the heart of Majorana based quantum information processing \cite{Sau10,Alicea11,Zhang13,Xue13,Knapp16,Aasen16,Zhang19,Wieckowski20}. Moreover, braiding and other logical gates could be effectively implemented by a sequence of parity measurements \cite{Bonderson08,Zheng16,Bomantara20,Knapp20,Karzig19,Zeng20}. Therefore, reading out the parity of MBSs is vital for both verifying the existence of MBSs and building a robust quantum computer. \\
\indent Practically, reading out the parity of two MBSs \cite{Hassler10,Hassler11,Gharavi16,Plugge17,Karzig17,Hoffman16,Hoffman17,Grimsmo19,Manousakis20,Szechenyi20,Steiner20,Munk20} is considered to be difficult. The reason is that the information of the parity is encoded in a pair of nonlocal MBSs. The generic method to tackle this problem is constructing an interference loop to map the nonlocal information to a local system. Usually, quantum dot system is employed to connect two or more MBSs \cite{Flensberg11,Bonderson11,Leijnse11,Jong19,Veen19,Zazunov20}. A natural idea to read out MBSs is parity-to-charge conversion \cite{Plugge17,Karzig17,Szechenyi20,Steiner20,Munk20}, which could benefit from the well-developed charge sensing technique in quantum dot system. For an example, Ref. \cite{Plugge17} employs the parity-dependent period of Rabi oscillation of double quantum dots to accomplish the conversion. This proposal requires fine-tuning the oscillation time to a specific value and is also very sensitive to other parameters. Instead, Refs. \cite{Karzig17} suggest to probe the parity through detecting the average dot occupation or the differential capacitance of the ground state of the quantum dot coupled with the Majorana island. Most recently,  the state evolution of the joint system under the charge sensing of the quantum dot is investigated \cite{Steiner20,Munk20}, and the susceptibilities of the readout schemes to low frequency charge fluctuation are analyzed~\cite{Maman20,Khindanov20}.\\
\begin{figure}
	\includegraphics[width=7cm]{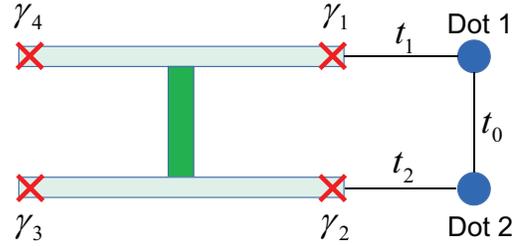}
	\caption{Majorana island coupled with DQD system. DQD consists of Dot 1 and 2 with inter-dot tunneling rate $t_0$. The superconducting island host four MBSs at its ends. The right two MBSs $\gamma_1, \gamma_2$ are tunneling coupled to Dot 1 and Dot 2, respectively. \label{fig1}}
\end{figure}
\indent In this paper, we construct a dynamical process to realize parity-to-charge conversion with a system illustrated in Fig. 1. A Majarana island hosts four MBSs, which locates at the ends of the nanowires. The right two MBSs are tunneling coupled to two quantum dots respectively. An electron in one dot could tunnel into another dot either directly, or mediated by the Majorana island. The latter tunneling  induces an effective Majorana-Majorana hybridization which depends 
on the parity of the Majorana pair. Thus, the interference of the two tunneling paths cause the total tunneling rate sensitive to the parity of MBSs. To read out the Majorana qubit, we apply Landau-Zener transitions by sweeping the voltage bias of the double quantum dot (DQD). After the sweep, the parity is imprinted on the charge state of the DQD, which can be detected by a charge sensor. By fine tuning the coupling between the two dots, a single-shot projective measurement can be realized. In the absence of fine-tuning, we propose a sequence of weak measurements, which eventually project Majorana qubit to either parity and form the strong projective readout. Compared with former schemes, this readout protocol has three virtues. Firstly, it is a non-demotion measurement as the MBSs parity is a conserved quantity in the readout process. Secondly, the charge is detected when the DQD is biased at its idle point. This feature makes the result be free of measurement back-action and immune to low frequency charge fluctuation. Lastly, the proposed discrete weak measurements enable us to monitor the state of the Majorana qubit by charge detections, which could be useful for (topologically unprotected) quantum information processing. \\
\indent This paper is organised as follows. The DQD-Majorana island system and its Hamiltonian are introduced in Sec. \ref{sec2}. After that, the readout scheme using Landau-Zener transition is present: we first investigate the single-shot readout with the requirement of fine-tuning in Sec. \ref{sec3A}, and then study the weak measurements without fine-tuning in Sec. \ref{sec3B} and the measurement-induced wavefunction collapse in Sec. \ref{sec3C}. We analyse the effects of relaxation and fluctuation noise on the measurement in Sec. \ref{sec4}. At last, we give a conclusion in Sec. \ref{sec5}.
\section{Hamiltonian of DQD-Majorana island system}\label{sec2}
As shown in Fig. 1, the Majorana island consists of two parallel one-dimensional topological superconductors connected by an ordinary superconductor. Under certain conditions, four MBSs would emerge at the ends of the topological superconductors, denoted by $\hat{\gamma}_{1,2,3,4}$. The coupling between two MBSs is exponentially suppressed with their distance. The Hamiltonian of the island reads
\begin{equation}\label{Eq1}
\hat{H}_q=E_C(\hat{N}-N_g)^2,
\end{equation}
where $E_C$ is the charge energy of the island, $\hat{N}$ is the total charge and $N_g$ denotes the offset. Due to the large $E_C$, the whole parity of the four MBSs is protected from quasiparticle tunneling between the island and its surrounding. Henceforth, we take the even subspace for instance. The two-fold degenerate states $\{|00\rangle, |11\rangle\}$ act as the basis vectors of a qubit, named topological qubit. \\
\indent Reading out the state of the topological qubit can be achieved by measuring the parity $p_{12}$ of modes $\gamma_1,\gamma_2$. To this end, $\gamma_1,\gamma_2$ are tunneling coupled to quantum dot 1 and 2, respectively. Meanwhile, the two quantum dots are bridged through a normal conductor to form a DQD. We assume that each quantum dot has a single nondegenerate spin-polarized state. Then, the Hamiltonian of DQD can be written as
\begin{equation}\label{Eq2}
H_{DQD}=\epsilon_1\hat{n}_1+\epsilon_2\hat{n}_2+t_0d_1^\dagger d_2+\text{h.c.}
\end{equation}
where $\hat{d}_i$ denotes the annihilation operator of the tunable level of dot $i$, $\hat{n}_i=\hat{d}_i^\dagger\hat{d}_i$ is the occupation. $\epsilon_i$ is the level energy of dot $i$, and $t_0$ is the hopping amplitude between dot 1 and 2 when they are resonant. Here, we assume that the two dots are spin polarized by the magnetic field, and the dot charge energy is the largest energy scale of the system. The island-DQD tunneling Hamiltonian could be expressed as
\begin{equation}\label{Eq3}
H_T=(t_1\hat{\gamma}_1\hat{d}_1+t_2\hat{\gamma}_2\hat{d}_2)e^{i\hat{\phi}/2}+\text{h.c.}
\end{equation}
where $t_i$ is the tunneling rate between mode $\hat{\gamma}_i$ and dot $i$, and $\hat{\phi}$ is the phase of the Majorana island with $[\hat{\phi}, \hat{N}]=2$.  In the measurement process, Majorana modes and quantum dots are largely detuned, thus there is no actual electron transferring between them. We consider the single electron subspace of the DQD, which is spanned by $\{|01\rangle_d, |10\rangle_d\}$. According to the perturbation theory, the effective Hamiltonian of the whole system read
\begin{equation}\label{Eq4}
H=H_q+\epsilon\sigma_z/2+(t_0+it_1t_2^*p_{12}/E_C)\sigma_x,
\end{equation}
where $\epsilon=\epsilon_1-\epsilon_2$ is the detuning of the DQD, and  $\sigma_z=|01\rangle_d\langle01|-|10\rangle_d\langle10|$, $\sigma_x=|01\rangle_d\langle10|+|10\rangle_d\langle01|$ are the Pauli operators.
\begin{figure}
	\includegraphics[width=7.5cm]{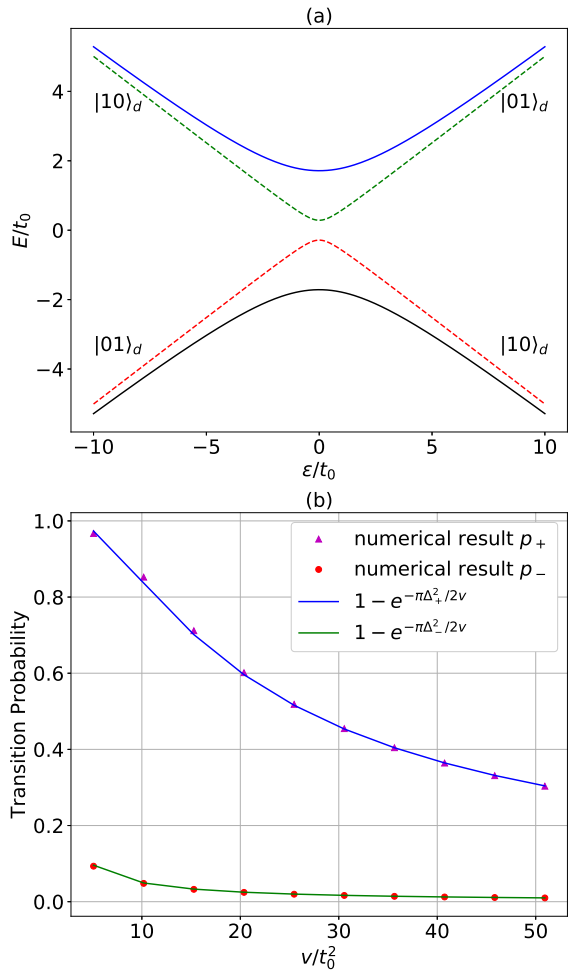}
	\caption{Energy levels and transition probability with $v$. (a). anticrossing structure of the enery levels of DQD conditioned on the parity $p_{12}=-1$(dashed lines), +1(solid lines). (b). transition probability from $|01\rangle_d$ to $|10\rangle_d$ with sweeping velocity $v$. The parameters are chosen as follows: $t_1=5t_0$, $t_2=it_1$, $E_C=35t_0$. \label{fig2}}
\end{figure}
\section{readout of MBSs via Landau-Zener transition}
Now we address the question how to read out the topological qubit using Landau-Zener transition. The energy levels are illustrated in Fig. 2a. The positive and negative eigenenergies of the DQD with $\epsilon$ shows up a anticrossing at $\epsilon=0$. Most importantly, the minimum gap of the anticrossing relies on the parity $p_{12}$:
\begin{equation}
	 \Delta_{\pm}=2(t_0\pm it_1t_2^*/E_C).
\end{equation}
\indent The ratio $\Delta_+/\Delta_-$ is the essential factor that determines the efficiency and fidelity of the measurement of the topological qubit. We assume $t_0, t_1$ are real and $|t_1|=|t_2|$ for simplicity (both assumptions can be loosed). We denote $t_2=t_1e^{i\varphi}$. As $\varphi=0$ or $\pi$, which leads to $|\Delta_+/\Delta_-|=1$, it is unable to discriminate the two parity states. On the contrary, if $\varphi=\pm \frac{\pi}{2}$, the gap bifurcates maximumly, and favour reading out the parity $p_{12}$. In fact, the phase difference $\varphi$ between $t_1$ and $t_2$ could be tuned in some manner, such as changing the magnetic flux of the loop formed by the superconductor and the DQD. Henceforth, we focus on the optimal case, and assume $t_2=i t_1$, which gives $\Delta_{\pm}=2(t_0\pm t_1^2/E_C)$. In order to read out the parity non-destructively, we should map it to an observable that is conserved during the measurement. 

\subsection{single-shot projective readouts of MBSs }\label{sec3A}
Single-shot projective measurements of MBSs could be realized if the parameters $t_0$ and $t_1$ were fine tuned to meet the relation $t_0=t_1^2/E_C$. In this case, the gap $\Delta_-$ for $p_{12}=-1$ is vanishing while $\Delta_+$ for $p_{12}=+1$ is finite. For simplicity, the parity states of $\gamma_1,\gamma_2$ are denoted as $|0\rangle_m(p_{12}=+1)$ and $|1\rangle_m(p_{12}=-1)$. The basic readout protocol is as follows. At the beginning, the DQD is set far away from the anticrossing point, and stay at the ground state $|01\rangle_d$. Sweep the detuning $\epsilon$ from a large negative value to the opposite side so that the DQD passes the anticrossing. After the sweep, the probability of the excited state of the DQD is related with the parity of the MBSs. If the sweep is adiabatic relative to the gap $\Delta_+$, the topological qubit would be entangled with the DQD, i.e.,  
\begin{equation*}
 (a|0\rangle_m+b|1\rangle_m)|01\rangle_d\rightarrow a|0\rangle_m|10\rangle_d+b|1\rangle_m|01\rangle_d.
\end{equation*}
At the end, we apply the charge sensor, e.g. a quantum point contact, to measure the state of the DQD; and the system collapse to one component of the entangled state according to the result obtained. Consequently, we could read off the parity of the MBSs.\\
\indent There are two remarkable merits of the above measurement approach. Firstly, it is a non-demolition measurement. That means the total parity of Majorana island is fixed after the readout process due to the island charge energy; and therefore, the charge states of the DQD are energy eigenstates as they are measured. Thus, it is allowed to repeat the measurement to enhance the signal noise ratio. Secondly, during the measurement process, the Majorana island stays at the ground state, therefore, the parity of MBSs is protected by Coulomb blockage from quasiparticle poisoning due to the large $E_C$. However, this single-shot readout requires challenging fine-tuning in experiment. The next subsection is devoted to put forward a measurement scheme without the fine-tuning.

\subsection{repeated weak measurements of MBSs}\label{sec3B}
Generally, both of the gaps $\Delta_{\pm}$ are non-vanishing. In this situation, the adiabatic sweeping always reaches the same DQD state $|10\rangle_d$ for both even and odd Majorana parities; the non-adiabatic sweeping will reach a superposition of the two DQD states. Therefore, the single sweep-and-detection protocol is unable to map the parity of the topological qubit to the DQD states. Thus, it is necessary to modify our measurement protocol.\\
\indent  The tactic to address this problem is applying weak measurement. Note that $\Delta_-\neq0$ and $\Delta_-<\Delta_+$. Initiate the DQD to the ground state $|01\rangle_d$ and sweep its detuning with a velocity $v=\frac{d\epsilon}{dt}$ through the anticrossing. Then, the DQD would transit to the state $|10\rangle_d$ with probability $p_\pm$ for the parity $p_{12}=\pm1$. Under the adiabatic-impulse approximation \cite{Shevchenko10}, $p_\pm$ can be obtained by Landau-Zener formula
\begin{equation}\label{LZ}
p_{\pm}=1-e^{-\pi\Delta_{\pm}^2/2v}.
\end{equation}
\indent We calculate $p_\pm$ as a function of $v$ numerically and compare the results with the adiabatic-impulse approximation, as shown in Fig. 2b. It can be seen that $p_-<p_+$ for any $v$ due to $\Delta_-<\Delta_+$. Moreover, the two transition probabilities descend with the increasing of $v$. Thus, it is possible to make $p_-$ vanishing while $p_+$ is finite by selecting proper $v$. In other words, if the parity of the MBSs $p_{12}=-1$, the DQD still stay at $|01\rangle_d$. Alternatively, DQD will end with a superposition state of $|01\rangle_d$ and $|10\rangle_d$. From the view of measurement, a single run of sweep-and-detection implements a weak measurement on MBSs instead of projective measurement.  \\
\begin{figure}
	\includegraphics[width=7.5cm]{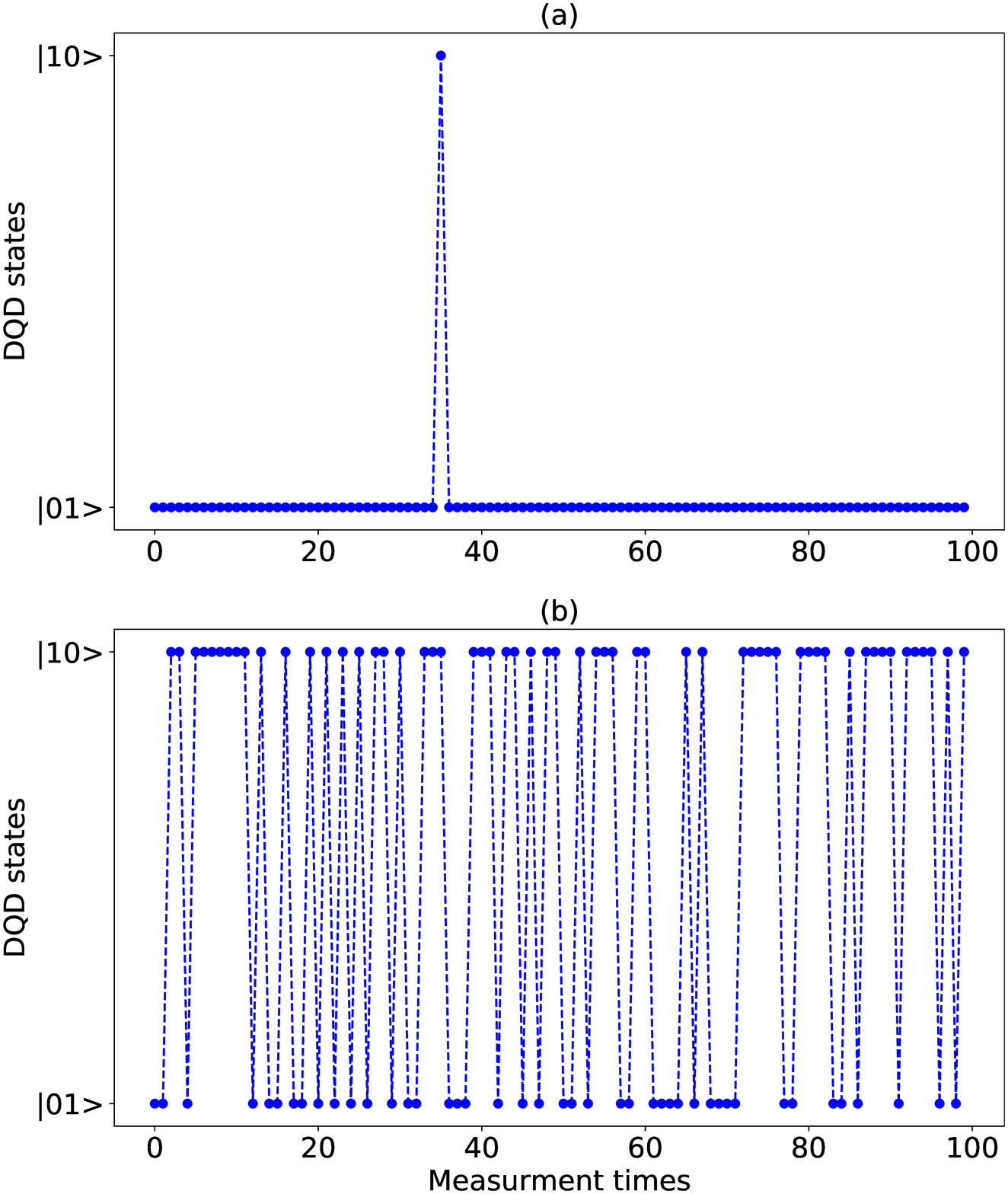}
	\caption{State trajectory of the DQD after each sweeping when the MBSs parity $p_{12}=-1$ (a) and $+1$ (b). The parameters are chosen as follows: $t_1=5t_0$, $t_2=it_1$, $E_C=35t_0$, the sweeping velocity is $v=80t_0^2/\pi$. Here, a sequence of one hundred runs of sweeping and detecting are simulated. Note that the state of the MBSs is unchanged in the process due to the non-demolition property of the measurement. \label{fig3}}
\end{figure}
In order to read out the MBSs, a sequence of repeated runs should be applied continuously. Note that before starting a new sweep, the DQD should be initiated to its ground state. We have simulated the state evolution trajectories of DQD for MBSs parity $-1$ and $+1$ respectively, as shown in Fig. 3. The whole process include 100 runs. It can be seen that when $p_{12}=-1$ there are barely charge transfers between the DQD. In contrast, the charge state of the DQD is obligated to jump back and forth frequently for the parity $+1$. Thus, the two parities of the MBSs could be clearly discriminated through a series of charge detections.

\subsection{wavefunction collapse of MBSs induced by weak measurements}\label{sec3C}
\indent Weak measurement will not project the target quantum system instantaneously to one eigenstate of the observable. Therefore, a superposition of the MBSs parity states would gradually collapse to one specific parity as repeated sweep-and-detection runs are applied. According to quantum trajectory theory, the state evolution of MBSs under weak measurement is stochastic and conditioned on the results of charge sensing.\\ 
\begin{figure}
\includegraphics[width=8.5cm]{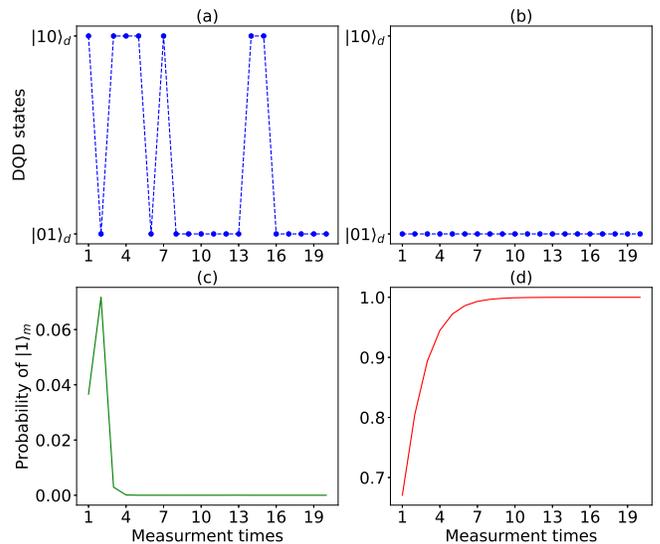}
\caption{State evolutions of the MBSs conditioned on the charge detection after each sweep. The initial state of MBSs $\gamma_1, \gamma_2$ is $(|0\rangle_m+|1\rangle_m)/\sqrt{2}$. The state trajectory of the DQD in (a) leads to the state of the MBSs collapse to $|0\rangle_m$, while the evolution of the DQD in (b) projects the MBSs to $|1\rangle_m$. The probabilities of $|1\rangle_m$ in these two processes are shown in (c) and (d), respectively.  \label{fig4}}	
\end{figure}
\indent There are two possible results of charge states of DQD after each sweep. No matter whether the charge state is changed or not during the sweep, the state of the MBSs has to be renewed due to the action of measurement. If the charge sensor observed a different charge distribution compared to that before the sweep, we know that the MBSs is probably with the parity $p_{12}=+1$. Therefore, the MBSs would collapse to this parity very sharply, which is similar to the so called quantum jump. Alternatively, if the DQD stayed at its initial state after the sweep, the state of the MBSs would be only gently modified because the detection gain very little information about the state. To reveal the relation between the detection results and the state trajectory of the MBSs, we have simulated the state evolution of the MBSs with an initial state  $(|0\rangle_m+|1\rangle_m)/\sqrt{2}$. Two typical state trajectories of the DQD and the MBSs are illustrated in Fig. 4.\\
\indent The first trajectory is shown in Fig. 4(a) and (c). From Fig. 4(a), we can see that the state of the DQD randomly jumps between $|01\rangle_d$ and $|10\rangle_d$. The repeat measurement precedure leads to a huge declining of the probability of $|1\rangle_m$ from 0.5 to 0.036, as depicted in Fig. 4(c). Although some oscillation occurs since then, another detection of $|10\rangle_d$ project the MBSs to the state $|0\rangle_m$ completely. On the contrary, in Fig. 4(b) the charge state of the DQD is maintained during the whole process of 20 runs of sweep-and-detection. In this case, the probability of $|1\rangle_m$  increases monotonously from 0.5 to 1 within ten sweeping periods, as shown in Fig. 4(d). From the two distinct trajectories, it can be seen that the initial state would totally collapse to either parity state after ten runs. That means we have applied a projective measurement on the parity state of the MBSs. Therefore, by measuring an ensemble of Majorana pairs with the same initial state, we could obtain the population of each parity in the state before the measurement.\\
\indent Before ending this subsection, we address how to calibrate the relevant parameters to gain an efficient readout. In fact, the overall measurement time is mainly determined by the least number of DQD sweeps needed to project the state of MBSs and the sweep duration. The sweep number can be minimized as $t_0\sim t_1^2/E_C$, which gives rise to a maximal ratio $\Delta_+/\Delta_-$. On the other hand, the DQD can be swept more quickly for larger $\Delta_+$ without loss of readout fidelity. In practice, we could tune the well-controlled quantities $t_0$ and the sweep velocity to find the optimal work point. To check it, we can apply continuously numerous runs of sweep-and-detection. If the charge state of DQD after each sweep distributes in two distinct manners, such as those shown in Fig. 3(a) and (b), the parameters are fitly calibrated. 

\section{effects of DQD relaxation and fluctuation noise}\label{sec4}
In our readout scheme, the state of the MBSs are partially entangled with that of the DQD through parity-dependent Landau-Zener processes. Thus, the ratio $p_+/p_-$ is the key quantity that determines the fidelity of the measurement. To get a high fidelity readout, the following conditions should be met: very small $p_-\ll1$ and a non-vanishing $p_+$. According to Eq. (\ref{LZ}), these conditions require the sweep velocity and the anticrossing gaps to satisfy the relations
\begin{equation} 
\pi\Delta_+^2/2v\sim 1, \pi\Delta_-^2/2v\ll 1.
\end{equation} 
Note that when $t_0\sim t_1^2/E_C$, the gap $\Delta_+$ is much larger than $\Delta_-$, which is benefit for mapping out the parity of MBSs. Under this situation, we can obtain distinct transition probabilities $p_\pm$ by adopting a suitable sweep velocity. Actually, the small $p_-$ may lead to unwanted detection results, such as the occasional jump in Fig. 3(a). However, this would not cause a measurement error if $p_+/p_-\gg1$, since our measurement consists of a sequence of sweep-and-detection. Thus, our scheme could achieve a high fidelity readout. However, the effects of decoherences have yet to be counted for. Now we address the question whether and to what extent the relaxation and fluctuation noise of DQD affect the fidelity of the measurement.
\subsection{DQD relaxation}
\indent The parity state of MBSs are partially mapped to the ground state and the excited state of DQD via Landau-Zener process. The fidelity of the map relies on the ratio of the transition probability, i.e., $p_+/p_-$. In practise, the relaxation effect of the DQD may modify the probabilities $p_\pm$, and accordingly reduce the visibility of the measurement results. The manner in which the relaxation affects the rates is to be clarified. For simplicity, here we account for the effect of relaxation with a constant characteristic time. Remember that the DQD is prepared at its ground state before each sweep, and the excitation probably occurs near the anticrossing. Hence, it is reasonable to believe that the relaxation has little effect on the readout fidelity if an half of the sweep duration was much shorter than the relaxation time. \\
\indent Now we discuss how to weaken the disturbance of the relaxation. Beside promoting the relaxation time of DQD, we could manage to minimize the duration of each sweep, which can be achieved by increasing the sweep velocity. However, on other side, the reduction of the sweep duration would diminish the rate $p_+$, which causes the measurement weaker. In this case, to accomplish the projective readout, more repetitions of sweep-and-detection are required, which would prolong the whole measurement time. Therefore, the sweep duration should be properly fixed according to the relaxation time and the other parameters that determine $p_+$. In fact
, it can be seen from Eq. \ref{LZ} that a larger gap $\Delta_+$ allows for faster sweeps while keeping the rate $p_+$ invariant.\\
\indent A viable parameter structure selected in this paper is as follows: $t_1=5t_0$, $t_2=it_1$, $E_C=35t_0$, $v=80t_0^2/\pi$. In this case, the transition probabilities are $p_+=0.51$, $p_-=0.02$, and the sweep duration is in the order of $2\pi/t_0$. On one hand, the ratio $p_+/p_-\sim 20$ is large enough to project completely the state of MBSs to either parity within several runs of sweep-an-detection, as shown in Fig. 4. On the other hand, $t_0$ could be tuned at will in the order of $100\text{MHz}\times2\pi$, which gives the single sweep duration about 10ns. Thus, the half length of each sweeping process is about 5ns, which is short than the relaxation time of DQD \cite{Petta04,Petersson10}. It is worth to note that the sweep duration can be further suppressed by optimizing the system parameters. Therefore, the measurement would not be destroyed by DQD relaxation. In addition, before each sweep, the energy decay of DQD is employed to initiate it to the ground state.
\begin{figure}
	\includegraphics[width=8cm]{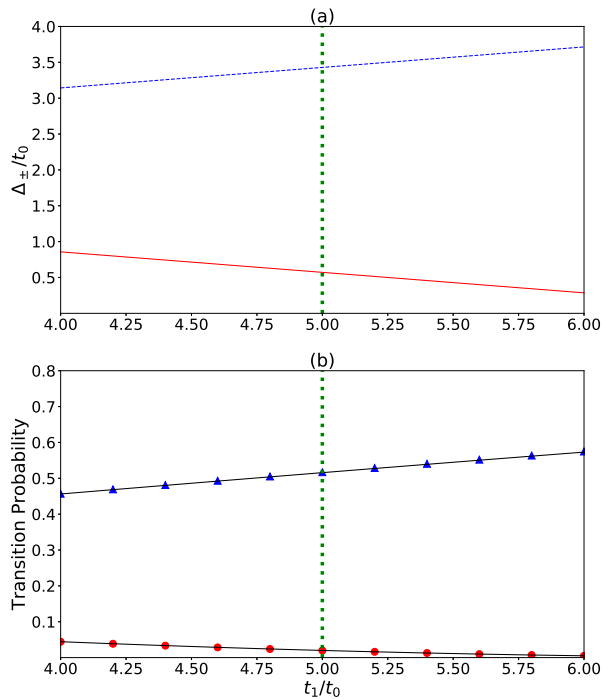}
	\caption{Energy gaps and transition probability with $t_1$. (a). Energy gaps of DQD conditioned on the parity $p_{12}=-1$(solid line), +1(dashed line). (b). The transition probability with tunneling rate $t_1$. The quantity is calculated both numerically (triangle for $p_{12}=+1$, closed circle for $p_{12}=-1$) and analytically (solid lines). The parameters are chosen as follows: $t_2=5it_0$, $v=80t_0^2/\pi$, $E_C=35t_0$. \label{fig5}}
\end{figure}
\subsection{Fluctuation noise}  
Generally, DQD is subject to two kinds of fluctuations: detuning fluctuation and tunneling fluctuation \cite{Petta04}. These fluctuations are stemmed from gate voltage noises, and usually play a part in low frequency of $1/f$ power spectrum \cite{Petersson10}. For DQD as a qubit, the detuning fluctuation is the dominant dephasing source. However, in our readout scheme with DQD-MBSs system, the detuning fluctuation has little influence on the measurement fidelity. The reason is that the DQD is swept from the large positive detuning to the opposite large detuning, instead of being statically biased. During each sweep, the fluctuation of the detuning $\epsilon$ is unable to modify the Landau-Zener transition probability \cite{Saito07}. At the stage of charge sensing, the detuning fluctuation would not affect the result either, because DQD is biased at the idle point. Therefore, the detuning fluctuation would not disturb the readout process. \\
\indent The fluctuations of tunnelings $t_i(i=0,1,2)$ would lead to variations of the gaps $\Delta_{\pm}$. Therefore, this type of fluctuations may reduce the measurement fidelity or efficiency. Due to the similar roles of these tunnelings on determining the gaps $\Delta_\pm$, we fix the value of $t_{0}$ and $t_2$, and investigate the influence of the variation of $t_1$ on the gaps and transition probability. As shown in Fig. 5a, the gaps vary in different trends with $t_1$. Under the assumption $\Delta_\pm>0$, the ratio $\Delta_+/\Delta_-$ monotonically rises with the increasing of $t_1$. Note that the vertical dotted line is drawn at $t_1=5\pi$ which is used in the former figures. The variations of the transition probability are shown in Fig. 5b. The sweep velocity is the same as that in Fig. 4. Remember that the readout fidelity relies on the ratio $p_+/p_-$. While $t_1$ fluctuates from $4t_0$ to $6t_0$, this ratio varies from about 10 to 110. On one side, a reduction of the ratio would blur the boundary between the two parities of MBSs; on the other side, an enhanced ratio would promote the readout fidelity. In our readout scheme, about ten sweep runs are needed to definitely project the state of MBSs. Thus, if $t_1$ distributed symmetrically around $5t_0$ during the measurement process, the statistical average of $p_+/p_-$ would approach the ratio at $t_1=5t_0$. Therefore, the tunneling fluctuation would not effectively contaminate the readout results if the fluctuation is not too large. 
\section{conclusion} \label{sec5} 
In summary, we put forward a protocol to read out the parity of MBSs employing Landau-Zener transition in Majorana island-DQD system. Through sweeping the bias of the DQD from large negative detuning to the opposite side, the parity is projected to the charge state of the DQD, which could be probed by charge sensing. This readout process enable us to uncover how the state of the MBSs collapses step by step based on the detection results. In addition, this readout is a non-demolition measurement, and behaves robust against the low frequency fluctuations. Therefore, it is feasible in experiment and could be applied to the quantum information processing and topological quantum computation.  
\section{Acknowledgments}
 ZTZ is funded by Introduction and Cultivation Plan of Youth Innovation Talents for Universities of Shandong Province (Research and Innovation Team on Materials Modification and Optoelectronic Devices at extreme conditions) and the National Nature Science Foundation of China (No.11404156). DEL acknowledge the support from NSF-China (Grant No.11974198).






\begin{thebibliography}{999}
\bibitem{Kitaev01} A. Y. Kitaev, Phys.-Usp. \textbf{44}, 131(2001).
\bibitem{Ivanov01}D. A. Ivanov, Phys. Rev. Lett. \textbf{86}, 268 (2001).
\bibitem{Nayak08}C. Nayak, S. H. Simon, A. Stern, M. Freedman, and S. Das Sarma, Rev. Mod. Phys. \textbf{80}, 1083 (2008).	
\bibitem{Akhmerov10}A. R. Akhmerov, Phys. Rev. B \textbf{82}, 020509 (2010)
\bibitem{Lutchyn10}R. M. Lutchyn, J. D. Sau, and S. Das Sarma, Phys. Rev. Lett. \textbf{105}, 077001 (2010).
\bibitem{Oreg10}Y. Oreg, G. Refael, and F. von Oppen, Phys.Rev.Lett. \textbf{105}, 177002 (2010).

\bibitem{Mourik12} V.Mourik, K.Zuo, S. M. Frolov, S.R.Plissard, E.P.A.M. Bakkers, and L. P. Kouwenhoven, Science \textbf{336}, 1003 (2012).
\bibitem{Das12} A. Das, Y. Ronen, Y. Most, Y. Oreg, M. Heiblum, and H. Shtrikman, Nat. Phys. \textbf{8}, 887 (2012).
\bibitem{Deng12}M. T. Deng, C. L. Yu, G. Y. Huang, M. Larsson, P. Caroff, and H. Q. Xu, Nano Lett. \textbf{12}, 6414 (2012).
\bibitem{Rokhinson12} L. P. Rokhinson, X. Liu, and J. K. Furdyna, Nat. Phys. \textbf{8}, 795 (2012).
\bibitem{Churchill13} H. O. H. Churchill, V. Fatemi, K. Grove-Rasmussen, M. T. Deng, P. Caroff, H. Q. Xu, and C. M. Marcus, Phys.Rev.B \textbf{87}, 241401(R) (2013).
\bibitem{Finck13}A. D. K. Finck, D. J. Van Harlingen, P. K. Mohseni, K. Jung, and X. Li, Phys. Rev. Lett. \textbf{110}, 126406 (2013).
\bibitem{Albrecht16}S. M. Albrecht, A. P. Higginbotham, M. Madsen, F. Kuemmeth, T. S. Jespersen, J. Nyg\r{a}rd, P. Krogstrup, and C. M. Marcus, Nature \textbf{531}, 206 (2016).
\bibitem{Deng16}M. T. Deng, S. Vaitiekenas, E. B. Hansen, J. Danon, M Leijnse, K. Flensberg, J. Nygård, P. Krogstrup, and C. M. Marcus, Science \textbf{354}, 1557 (2016).
\bibitem{Chen17}J. Chen, P. Yu, J. Stenger, M. Hocevar, D. Car, S. R. Plissard, E. P. A. M. Bakkers, T. D. Stanescu, and S. M. Frolov, Sci. Adv. \textbf{3}, e1701476 (2017).
\bibitem{Suominen17} H. J. Suominen, M. Kjaergaard, A. R. Hamilton, J. Shabani, C. J. Palmstrøm, C. M. Marcus, and F. Nichele, Phys. Rev. Lett. \textbf{119}, 176805 (2017).
\bibitem{Nichele17} F. Nichele, A. C. C. Drachmann, A. M. Whiticar, E. C. T. O’Farrell, H. J. Suominen, A. Fornieri, T. Wang, G. C. Gardner, C. Thomas, A. T. Hatke, P. Krogstrup, M. J. Manfra, K. Flensberg, and C. M. Marcus, Phys.Rev.Lett. \textbf{119}, 136803 (2017).
\bibitem{Onder18}\"{O}nder G\"{u}l, H. Zhang, J. D. S. Bommer, M. W. A. de Moor, D. Car, S. R. Plissard, E. P. A. M. Bakkers, A. Geresdi, K. Watanabe, T. Taniguchi et al., Nat. Nanotechnol. \textbf{13}, 192 (2018).
\bibitem{Sau10}J. D. Sau, S. Tewari, and S. D. Sarma, Phys. Rev. A \textbf{82}, 052322 (2010).
\bibitem{Alicea11} J. Alicea, Y. Oreg, G. Refael, F. von Oppen, and M. P. A. Fisher, Nat. Phys. \textbf{7}, 412 (2011).
\bibitem{Zhang13} Z.-T. Zhang and Y. Yu, Phys. Rev. A \textbf{87}, 032327 (2013).
\bibitem{Xue13} Z.-Y. Xue, L. B. Shao, Y. Hu, S.-L. Zhu, and Z. D. Wang, Phys. Rev. A \textbf{88}, 024303 (2013).
\bibitem{Knapp16} C. Knapp, M. Zaletel, D. E Liu, M. Cheng, P. Bonderson, and C. Nayak, Phys. Rev. X \textbf{6}, 041003 (2016).
\bibitem{Aasen16}D. Aasen, M. Hell, R. V. Mishmash, A. Higginbotham, J. Danon, M. Leijnse, T. S. Jespersen, J. A. Folk, C. M. Marcus, K. Flensberg, and J. Alicea, Phys. Rev. X \textbf{6}, 031016 (2016).
\bibitem{Zhang19}Z.-T. Zhang, F. Mei, X.-G. Meng, B.-L. Liang, and Z.-S. Yang, Phys. Rev. A \textbf{100}, 012324 (2019).
\bibitem{Wieckowski20} A. Wieckowski, M. Mierzejewski, and M. Kupczy\'{n}ski, Phys. Rev. B \textbf{101}, 014504 (2020).
\bibitem{Bonderson08} P. Bonderson, M. Freedman, and C. Nayak, Phys. Rev. Lett. \textbf{101}, 010501 (2008).
\bibitem{Zheng16}H. Zheng, A. Dua and L. Jiang, New J. Phys. \textbf{18}, 123027 (2016).
\bibitem{Bomantara20} R. W. Bomantara, and J. Gong,
Phys. Rev. B \textbf{101}, 085401 (2020).
\bibitem{Knapp20} C. Knapp, J. I. V\"{a}yrynen, and R. M. Lutchyn,
Phys. Rev. B \textbf{101}, 125108 (2020).
\bibitem{Karzig19}T. Karzig, Y. Oreg, G. Refael, and M. H. Freedman, Phys. Rev. B \textbf{99}, 144521 (2019).
\bibitem{Zeng20}C. Zeng, G. Sharma, T. D. Stanescu, and S. Tewari, Phys. Rev. B \textbf{102}, 205101 (2020).
\bibitem{Hassler10}F. Hassler, A. R. Akhmerov, C-Y. Hou, and C. W. J. Beenakker, New J. Phys. \textbf{12}, 125002 (2010).
\bibitem{Hassler11} F. Hassler, A. R. Akhmerov, and C. W. J. Beenakker, New J. Phys. \textbf{13}, 095004 (2011).
\bibitem{Gharavi16} K. Gharavi, D. Hoving, and J. Baugh, Phys. Rev. B \textbf{94}, 155417 (2016).
\bibitem{Grimsmo19} A. L. Grimsmo and T. B. Smith, Phys. Rev. B \textbf{99}, 235420 (2019).
\bibitem{Plugge17} S. Plugge, A. Rasmussen, R. Egger, and K. Flensberg, New J. Phys. \textbf{19}, 012001 (2017).
\bibitem{Karzig17} T. Karzig, C. Knapp, R. M. Lutchyn, P. Bonderson, M. B. Hastings, C. Nayak, J. Alicea, K. Flensberg, S. Plugge, Y. Oreg, C. M. Marcus, and M. H. Freedman, Phys. Rev. B \textbf{95}, 235305 (2017).
\bibitem{Hoffman16} S. Hoffman, C. Schrade, J. Klinovaja, and D. Loss, Phys. Rev. B \textbf{94}, 045316 (2016).
\bibitem{Hoffman17} S. Hoffman, D. Chevallier, D. Loss, and J. Klinovaja, Phys. Rev. B \textbf{96}, 045440 (2017). 
\bibitem{Szechenyi20}G. Sz\'{e}chenyi and A. P\'{a}lyi, Phys. Rev. B \textbf{101}, 235441 (2020).
\bibitem{Steiner20} J. F. Steiner and F. von Oppen, Phys. Rev. Research \textbf{2}, 033255 (2020).
\bibitem{Munk20} M. I. K. Munk, J. Schulenborg, R. Egger, and K. Flensberg, Phys. Rev. Research \textbf{2}, 033254 (2020).
\bibitem{Manousakis20} J. Manousakis, C. Wille, A. Altland, R. Egger, K. Flensberg, and F. Hassler, Phys. Rev. Lett. \textbf{124}, 096801 (2020).
\bibitem{Flensberg11} K. Flensberg, Phys. Rev. Lett. \textbf{106}, 090503 (2011).
\bibitem{Bonderson11} P. Bonderson and R. M. Lutchyn, Phys. Rev. Lett. \textbf{106}, 130505 (2011).
\bibitem{Leijnse11} M. Leijnse, and K. Flensberg, Phys. Rev. Lett. \textbf{107}, 210502 (2011).
\bibitem{Jong19}D. De Jong, J. Van Veen, L. Binci, A. Singh, P. Krogstrup, L. P. Kouwenhoven, W. Pfaff, and J. D. Watson, Phys. Rev. Applied \textbf{11}, 044061 (2019).
\bibitem{Veen19}J. van Veen, D. de Jong, L. Han, C. Prosko, P. Krogstrup,
J. D. Watson, L. P. Kouwenhoven, and W. Pfaff, Phys. Rev. B \textbf{100}, 174508 (2019).
\bibitem{Zazunov20}A. Zazunov, R. Egger, and Y. Gefen,
Phys. Rev. Research \textbf{2}, 023054 (2020).
\bibitem{Maman20}V. D. Maman, M. Gonzalez-Zalba, and A. P\'{a}lyi, arXiv:2006.12391 (2020).
\bibitem{Khindanov20} A. Khindanov, D. Pikulin and T. Karzig, arXiv:2007.11024 (2020).
\bibitem{Shevchenko10}S. Shevchenko, S. Ashhab, and F. Nori, Phys. Rep. \textbf{492}, 1 (2010).
\bibitem{Petta04}J. R. Petta, A. C. Johnson, C. M. Marcus, M. P. Hanson, and A. C. Gossard, Phys. Rev. Lett. \textbf{93}, 186802 (2004).
\bibitem{Petersson10} K. D. Petersson, J. R. Petta, H. Lu, and A. C. Gossard
Phys. Rev. Lett. \textbf{105}, 246804 (2010).
\bibitem{Saito07} K. Saito, M. Wubs, S. Kohler, Y. Kayanuma, and P. H\"{a}nggi, Phys. Rev. B \textbf{75}, 214308 (2007).
\end{thebibliography}

\end{document}